\documentclass[aps,prb,amsmath,amssymb,twocolumn]{revtex4}
\usepackage{graphicx}
\usepackage{hyperref}

\begin{document}

\title{Spin-current quantization in a quantum point contact with spin-orbit interaction}

\author{Vladimir~A.~Sablikov}

\affiliation{Kotel'nikov Institute of Radio Engineering and Electronics,
Russian Academy of Sciences, Fryazino, Moscow District, 141190,
Russia}

\begin{abstract}
We develop a realistic and analytically tractable model to describe the spin current which arises in a quantum point contact (QPC) with spin-orbit interaction (SOI) upon a small voltage is applied.  In the model, the QPC is considered as a saddle point of two-dimensional potential landscape. The SOI acts within a finite region and is absent deep in the reservoirs. The SOI strength is not supposed to be strong. It is shown that the spin polarization appears in the third order of the perturbation theory as a result of definite combinations of electron transitions. They include two intersubband transitions to nearest subbands and one intrasubband transition. The spin current is proportional to the cube of the SOI strength and strongly depends on geometric parameters of the saddle point. The spin is polarized in the plane of the QPC and directed normally to the electron current if the SOI is of Rashba type. As a function of the saddle-point potential (i.e., the height of the QPC barrier), the spin conductance and especially the spin polarization have characteristic features (specifically, peaks) correlated with the charge conductance quantization steps. The peak shape depends on the length of the region where the SOI acts. In QPCs with sharp potential landscape, this picture is distorted by interference processes.
\end{abstract}
\maketitle

\section{Introduction}

Spin transport in quantum point contacts (QPCs) with spin-orbit interaction (SOI) attracts much interest because of non-trivial spin dynamics in a laterally confined electron system connected to electronic reservoirs. The interest is stimulated also by direct observations of the spin-polarized electron current passing through the QPC in the presence of SOI (Ref.~\onlinecite{Rokhinson}) and recent experiments in which spin-filtering properties of QPCs were successfully used to manipulate the electron spin in nonmagnetic systems based on two-dimensional electron gas in GaAs.~\cite{Koop,Frolov}.

A signature of QPCs is the conductance quantization staircase that occurs in the systems when electrons move ballistically through independent conducting channels (transverse quantization subbands). The conductance is well described by famous Landauer formula~\cite{Landauer}
\begin{equation}
 G=\dfrac{2e^2}{h}\sum\limits_n^N T_n\,,
\label{conductance}
\end{equation} 
where $T_n$ is the transmission coefficient for $n$th conducting channel not including spin. With decreasing the barrier height $U_0$ in the QPC, the channels are successively opened giving rise to a steplike increase in the conductance $G$. This effect was observed in many experiments.~\cite{experiment}

In the presence of SOI the situation becomes much more complicated. When passing through the QPC the electron flow acquires spin polarization, and therefore, spin conductance should be considered along with the charge conductance. A rather general consideration of the charge and spin transport in two-terminal ballistic structures with SOI is based on the scattering matrix formalism.~\cite{Buttiker_S} In this way Landauer formulas for the charge and spin conductances were derived for a mesoscopic system coupled to electron reservoirs in which the SOI vanishes.~\cite{ZhaiPRL,Entin-Wohlman} Since the two conductances are generated by unique scattering matrix, they should correlate~\cite{Entin-Wohlman}. The ascertainment of such a correlation is obviously very important for the investigation of the spin transport since it enables one to identify correctly the spin-polarization mechanism in experiments. Ref.~\onlinecite{Entin-Wohlman} demonstrated the correlations between the spin and charge conductances and their quantization for transport through a cylinder with the SOI acting in a stripe.

In the case of QPCs the problem of the spin conductance quantization remains insufficiently investigated yet, though the QPCs are one of most important model systems in mesoscopic physics. There is no unique opinion about specific features of the spin conductance, considered as a function of the barrier height or the Fermi energy, and their relation to the charge conductance features.

Moreover, the spin-polarization mechanism in QPCs is not well understood, though it is clear from recent works that two factors are important: intersubband transitions of electrons~\cite{Governale,ZhaiPRL,Eto} and the presence of transition regions where a laterally confined part of the QPC (quasi-one-dimensional channel) widens to the size of electron reservoirs of higher dimensionality~\cite{Eto,Silvestrov}. Particularly, unusual trajectories were revealed by Silvestrov and Mishchenko~\cite{Silvestrov} within the quasi-classical approach to exist in these regions. Eto et al.~\cite{Eto,Yokouchi} suggested that intersubband transitions, owing to which the polarization arises, occur just in the outer side of the QPC adjoining to anode reservoir. 

Numerical evaluations of the spin polarization are quite optimistic. They show that the polarization can exceed 50\% (Refs~\onlinecite{Eto,Yokouchi,Reynoso1,Reynoso2,Silvestrov}), but the estimated polarization is very different depending on the specific models and parameters of the QPC used by different authors in their numerical calculations. There is a lack of theoretical studies of the spin conductance dependences of on geometric parameters of the QPC, as well as on the barrier height in the QPC. Numerous calculations of electron waveguides with step-like constrictions in the presence of the SOI (Refs~\onlinecite{Zhai1,Liu,Zhai2}) do not clarify these questions since the results they give are very complicated because of strong interference effect.

The present work aims to clarify the mechanism of spin polarization and find out characteristic features of the spin conductance within a realistic model of a QPC with SOI, which would be consistent with experimental situation where the geometrical shape of the QPC is rather smooth. We develop an analytical theory by considering the QPC as a saddle point in two-dimensional potential landscape. This is a very reasonable model, which was successfully used~\cite{Glazman,Buttiker} to explain the conductance quantization steps in good agreement with the experiments. 

In this paper we generalize this model by including the SOI. In doing this we suppose that the SOI is localized in the vicinity of the QPC and is absent deep in the reservoirs. This is a natural assumption in the case where the SOI of the Rashba type is created by gates forming the QPC. Moreover, in such a way the ambiguity of the spin current definition in systems with SOI is avoided. The spin current is calculated in the reservoirs where it is well defined.~\cite{note1} To be specific we consider the spin current injected into the right reservoir when electrons driven by an applied voltage move from left to right. In addition, we restrict ourselves by the case where the SOI is weak, more precisely, the characteristic energy of the SOI is small compared with other characteristic energies (such as the barrier height, the Fermi energy and the intersubband energy). This assumption is well justified for the structures fabricated on the basis of such semiconductors as InAs and GaAs, which are used in present experiments with ballistic QPCs.

In the frame of this model, the charge conductance practically coincides with the standard quantization staircase since the SOI induced correction to Eq.~(\ref{conductance}) is of the second order in the SOI strength. The spin current arises in the third order of the perturbation theory. The spin conductance is determined by definite combinations of three matrix elements of electron transitions, two of which are intersubband transitions and one is intrasubband transition. We take into account all possible transitions to find finally the spin conductance and analyze its dependence on the parameters of the saddle-point potential and the spatial distribution of the SOI strength. In the case of the Rashba SOI, the spin current is polarized in the plane of the QPC normally to the particle current.

The spin conductance magnitude is found to depend strongly on the longitudinal and transverse lengths of the saddle-point potential landscape. The spatial distribution of the SOI strength affects the form of the function describing the dependence of the spin conductance on the barrier height. A general feature of this function is the presence of a maximum and nearby inflection point, the position of which correlates with the charge conductance steps. In addition, we show how this feature is distorted by the interference effect in the QPCs with sharp potential landscape.

\section{Model and transmission matrix}

Consider a QPC as a constriction created by gates in two-dimensional electron gas. We approximate the potential landscape by the function $U(x,y)$ with a saddle point:
\begin{equation}
 U(x,y)=\frac{U_0}{\cosh^2(x/L)}+\frac{m^*\omega_y^2y^2}{2}\,,
\label{potential_landscape}
\end{equation}
where $x$ is the longitudinal coordinate, along which the current flows, $y$ is transverse coordinate, $m^*$ is effective mass of electrons. 

This model potential is attractive because it admits an exact analytical solution and well simulates the QPCs studied in experiments.

Suppose that the SOI acts in a region of finite size in the vicinity of the QPC and is absent deep inside the electronic reservoirs. When a small bias voltage $V$ is applied across the source and drain reservoirs, an unpolarized electron flow falls on the QPC from the left reservoir, acquires the spin polarization in the QPC and goes to the right reservoir.

For the sake of simplicity we consider the case where the SOI strength depends only on the coordinate $x$. In the case of the Rashba SOI, the SOI Hamiltonian reads:
\begin{equation}
 H_{so}=\frac{\alpha(x)}{\hbar}\left(p_y\sigma_x-p_x\sigma_y\right)+\frac{i}{2}\frac{d\alpha}{dx}\sigma_y\,,
\end{equation}
where $p_x$ and $p_y$ are electron momentum components, $\sigma_x$ and $\sigma_y$ are Pauli matrices, $\alpha(x)$ is a function describing the strength of the Rashba SOI. In what follows we assume that $\alpha(x)$ is an even function which goes to zero as $x\to \pm\infty$.

The Hamiltonian of the system is
\begin{equation}
 H=\frac{p_x^2+p_y^2}{2m^*}+U(x,y)+H_{so}\,
\end{equation}
We find the eigenfunctions of $H$ by considering $H_{so}$ as a perturbation. The eigenfunctions $\Psi$ are expressed via eigenfunctions of the unperturbed Hamiltonian $H_0=(p_x^2+p_y^2)/2m^*+U(x,y)$, which are well known in the literature.~\cite{Landau}

\subsection{Unperturbed states}
The unperturbed eigenstates are the product of longitudinal, transverse and spin functions 
\begin{equation}
 |rnks\rangle=|rk\rangle|n\rangle|s\rangle = \psi_{k}^{(r)}(x)\varphi_n(y)\chi_s\,.
\end{equation}

Here $|n\rangle$ is the transverse wave function
\begin{equation}
 \varphi_n(y)\!=\!\frac{1}{\pi^{1/4}\sqrt{w}}\frac{1}{\sqrt{2^nn!}}\exp{\left(\!-\frac{y^2}{2w^2}\right)}H_n\left(\!\frac{y}{w}\right),
\end{equation}
with $H_n$ being the Hermitian polynomial, $w=\sqrt{\hbar/m^*\omega_y}$ being the characteristic width of the QPC. The number $n=0,1,2,3,\cdots$ defines the subband energies:
\begin{equation}
 \varepsilon_n=\left(\!n+\frac{1}{2}\right)\hbar \omega_y\,.
\end{equation}

$|rk\rangle$ is the longitudinal wave function:
\begin{equation}
\begin{split}
 \psi_k^{(r)}(x)= & \frac{\Gamma(a)\Gamma(b)}{\Gamma(c)\Gamma(a+b-c)}\left[2\cosh\left(\frac{x}{L}\right)\right]^{ikL} \\
&\times {_2F_1}\left(a,b;c;\frac{1-r\tanh(x/L)}{2}\right)
\end{split}
\end{equation}
where $r=\pm$ stands for right- and left-moving waves incident on the QPC from the left and right reservoirs; $k$ is the wave vector defined as a positive value; ${_2F_1}(a,b,c;\xi)$ is the Gauss hyper-geometric function, with $a$, $b$, and $c$ being functions of $k$ 
\begin{align*}
 a(k)&=\frac{1}{2}-ikL+\sqrt{\frac{1}{4}-\frac{2m^*U_0L^2}{\hbar^2}}\\
 b(k)&=\frac{1}{2}-ikL+\sqrt{\frac{1}{4}+\frac{2m^*U_0L^2}{\hbar^2}}\\
 c(k)&=1-ikL\,.
\end{align*}

$|s\rangle$ is the spin eigenfunction of $\sigma_z$ matrix with eigenvalue $s=\pm 1$.

The energy eigenvalue is
\begin{equation}
 E_{n,k}=\varepsilon_n+ \frac{\hbar^2k^2}{2m^*}\,.
\end{equation}

Below we consider the spin current produced by an unpolarized electron flow incident on the QPC from the left reservoir with the energy at the Fermi level. Unperturbed wave functions of these electrons $|+nk_ns\rangle$ behave asymptotically at $x\to +\infty$ as
\begin{equation}
 |+nk_ns\rangle\simeq t_{k_n}\exp(ik_nx)\varphi_n(y)\chi_s\,,
\end{equation}
where
\begin{equation}
 k_n=\frac{\sqrt{2m^*(E-\varepsilon_n)}}{\hbar}\,,
\end{equation} 
$t_{k_n}$ is the transmission coefficient
\begin{equation}
 t_{k_n}= \frac{\Gamma[a(k_n)]\Gamma[b(k_n)]}{\Gamma[c(k_n)]\Gamma[a(k_n)+b(k_n)-c(k_n)]}\,.
\end{equation} 

\subsection{Transmission matrix}
Perturbed wave functions $|\Psi^{(r)}_{nk_ns}\rangle$ are calculated up to the third order in the SOI strength, since the spin current appears in the third order of the perturbation theory. We do not write out the total wave function and restrict ourselves by its asymptotic expression at $x\to\infty$, which is only needed to calculate the spin current generated in the QPC. The wave function of right-moving electrons is presented via the transmission matrix $t_{n_1s_1,ns}$
\begin{equation}
 |\Psi^{(+)}_{nk_ns}\rangle \Big|_{x\to\infty}\simeq \sum\limits_{n_1,s_1}t_{n_1s_1,ns}e^{ik_{n_1}x}\varphi_{n_1}(y)|s_1 \rangle\,.
\end{equation} 
The transmission matrix components have the following form:
\begin{widetext}
\begin{equation}
t_{n_1s_1,ns}\!=\!t_{k_{n_1}}\!\left\{\!\left[\!\delta_{n_1n}\!+\!\frac{im^*}{\hbar^2k_{n_1}}\left(sG^{(2)}_{n_1n}\!+\!H^{(2)}_{n_1n}\!+\!\cdots\!\right)\right]\!\delta_{s_1s}\!
+\!\frac{im^*}{\hbar^2k_{n_1}}\left[\!s\!\left(G^{(1)}_{n_1n}\!+\!G^{(3)}_{n_1n}\!+\!\cdots\!\right)\!+\!H^{(1)}_{n_1n}\!+\!H^{(3)}_{n_1n}\!+\!\cdots\!\right]\!\delta_{s_1\bar{s}}\!\right\},
\label{trasmis_matrix}
\end{equation} 
where $\bar{s}=-s$,
\begin{equation}
 G^{(1)}_{mn}=\delta_{mn}F^x_{+k_n,+k_n}\,,\quad H^{(1)}_{mn}=if_{mn}F^y_{+k_m,+k_n}\,,
\end{equation} 
\begin{equation}
 G^{(2)}_{mn}=if_{mn}\sum\limits_r^{\pm}\int\frac{dk'}{2\pi}\left[\frac{F^x_{+k_m,rk'} F^y_{rk',+k_n}}{E-\varepsilon_m(k')+i0}-\frac{\sum_r^{\pm} F^y_{+k_m,rk'} F^x_{rk',+k_n}}{E-\varepsilon_n(k')+i0}
\right]\,,
\end{equation} 
\begin{equation}
 H^{(2)}_{mn}=\delta_{mn}\sum\limits_r^{\pm}\int\frac{dk'}{2\pi}\frac{F^x_{+k_m,rk'} F^x_{rk',+k_n}}{E-\varepsilon_n(k')+i0}+ \sum\limits_{m'}f_{mm'}f_{m'n}\sum\limits_r^{\pm}\int\frac{dk'}{2\pi}\frac{F^y_{+k_m,rk'} F^y_{rk',+k_n}}{E-\varepsilon_{m'}(k')+i0},
\end{equation} 
\begin{equation}
\begin{split}
G^{(3)}_{mn}= &-\delta_{mn}\sum\limits_{r_1r_2}^{\pm}\iint\frac{dk_1dk_2}{(2\pi)^2}\frac{F^x_{+k_m,r_1k_1} F^x_{r_1k_1,r_2k_2} F^x_{r_2k_2,r+k_n}}{[E-\varepsilon_n(k_1)+i0][E-\varepsilon_n(k_2)+i0]}\\
&-\sum\limits_{m_1}f_{mm_1}f_{m_1n}\sum\limits_{r_1r_2}^{\pm}\iint\frac{dk_1dk_2}{(2\pi)^2}\left\{\frac{F^y_{+k_m,r_1k_1} F^y_{r_1k_1,r_2k_2} F^x_{r_2k_2,+k_n}}{[E-\varepsilon_{m_1}(k_1)+i0][E-\varepsilon_{n}(k_2)+i0]} \right.\\
&\left.+\frac{F^x_{+k_m,r_1k_1} F^y_{r_1k_1,r_2k_2} F^y_{r_2k_2,+k_n}}{[E-\varepsilon_{m}(k_1)+i0][E-\varepsilon_{m_1}(k_2)+i0]}-\frac{F^y_{+k_m,r_1k_1} F^x_{r_1k_1,r_2k_2} F^y_{r_2k_2,+k_n}}{[E-\varepsilon_{m_1}(k_1)+i0][E-\varepsilon_{m_1}(k_2)+i0]}\right\},
\end{split}
\end{equation} 
\begin{equation}
\begin{split}
H^{(3)}_{mn}&=-if_{mn}\sum\limits_{r_1r_2}^{\pm}\iint\frac{dk_1dk_2}{(2\pi)^2} \left[\frac{F^x_{+k_m,r_1k_1}F^x_{r_1k_1,r_2k_2}F^y_{r_2k_2,+k_n}}{[E\!-\!\varepsilon_m(k_1)\!+\!i0][E\!-\!\varepsilon_m(k_2)\!+\!i0]}-\frac{F^x_{+k_m,r_1k_1}F^y_{r_1k_1,r_2k_2}F^x_{r_2k_2,+k_n}}{[E\!-\!\varepsilon_m(k_1)\!+\!i0][E\!-\!\varepsilon_n(k_2)\!+\!i0]}\right.\\
&\left.+\frac{F^y_{+k_m,r_1k_1}F^x_{r_1k_1,r_2k_2}F^x_{r_2k_2,+k_n}}{[E\!-\!\varepsilon_n(k_1)\!+\!i0][E\!-\!\varepsilon_n(k_2)\!+\!i0]}\right]\!-i\!\!\sum\limits_{m_1m_2}\!f_{mm_1}f_{m_1m_2}f_{m_2n}\sum\limits_{r_1r_2}^{\pm} \iint\!\frac{dk_1dk_2}{(2\pi)^2}\frac{F^y_{+k_m,r_1k_1}F^y_{r_1k_1,r_2k_2} F^y_{r_2k_2,+k_n}}{[E\!-\!\varepsilon_{m_1}(k_1)\!+\!i0][E\!-\!\varepsilon_{m_2}(k_2)\!+\!i0]}.
\end{split}
\end{equation} 
Here $F^{x,y}_{r_1k_1,r_2k_2}$ and $f_{mn}$ are longitudinal and transverse components of the matrix element of the SOI Hamiltonian
\begin{equation}
 \langle r_1n_1k_1s_1|H_{so}|r_2n_2k_2s_2\rangle=-\delta_{s_1\bar{s}_2}\left(s_1F^x_{r_1k_1,r_2k_2}\delta_{n_1n_2}+iF^y_{r_1k_1,r_2k_2}f_{n_1n_2}\right),
\end{equation} 
\end{widetext}
\begin{equation}
 \begin{array}{ll}
 F^x_{r_1k_1,r_2k_2}&=\langle r_1k_1|\alpha\dfrac{d}{dx}+\dfrac{1}{2}\dfrac{d\alpha}{dx}|r_2k_2\rangle\\ 
 F^y_{r_1k_1,r_2k_2}&=\langle r_1k_1|\alpha|r_2k_2\rangle\,,  ×
 \end{array}
\label{F_matrix}
\end{equation} 
\begin{equation}
 f_{mn}=\langle m|\dfrac{d}{dy}|n\rangle\,.
\label{fmn_matrix} 
\end{equation} 
$F^{x,y}_{r_1k_1,r_2k_2}$ and $f_{mn}$ satisfy the following symmetry relations:
\begin{equation}
\begin{array}{ll}
 &F^x_{r_1k_1,r_2k_2}\!=\!-\!(F^x_{r_2k_2,r_1k_1})^*,\\ &F^x_{r_1k_1,r_2k_2}\!=\!-\!F^x_{\bar{r}_1k_1,\bar{r}_2k_2},\\
 &F^y_{r_1k_1,r_2k_2}\!=\!(F^y_{r_2k_2,r_1k_1})^*,\\ &F^y_{r_1k_1,r_2k_2}\!=\!F^y_{\bar{r}_1k_1,\bar{r}_2k_2},
\end{array}
\label{Fxy_symmetry}
\end{equation}
\begin{equation}
 f_{mn}=-f_{nm}\,.
\label{fmn_symmetry}
\end{equation} 
Specifically in the case of the parabolic confining potential in $y$ direction, $f_{mn}$ matrix simplifies to 
\begin{equation}
 f_{mn}=\frac{1}{w\sqrt{2}}\left(\delta_{m,n-1}\sqrt{n}-\delta_{m,n+1}\sqrt{n+1}\right).
\label{fmn_parabolic}
\end{equation} 

The transmission matrix $t_{n_1s_1,ns}$ defined by Eq.~(\ref{trasmis_matrix}) satisfies the symmetry relations following from the time-reversal symmetry and the invariance with respect to $x$ and $y$ inversion.~\cite{ZhaiPRL} 

\section{Spin current}
The spin current is generated in the QPC by right-moving states in the energy layer $eV$, where $V$ is an applied voltage. The spin current in the right reservoir is defined as follows: 
\begin{equation}
 \mathcal{J}^{\beta}_s=eV\frac{\hbar}{2}\sum\limits_{ns}D_n(E)\int\limits_{-\infty}^{\infty}dy\Psi^{(+)^+}_{nk_ns}(v_n\hat{\sigma}_{\beta})\Psi^{(+)}_{nk_ns},
\end{equation} 
where $D_n(E)$ is the density of states in $n$-th subband, $v_n(E)$ is the velocity, $\beta$ denotes the spin projections.

The spin current has only one nonzero component $\mathcal{J}^y_s$ polarized in $y$ direction that is expressed in terms of the $t_{n_1s_1,ns}$ matrix,~\cite{ZhaiPRL}
\begin{equation}
 \mathcal{J}^y_s=\frac{eV}{4\pi}\sum\limits_{nm}\mathrm{Im}\left(t^*_{m\uparrow,n\uparrow}t_{m\downarrow,n\uparrow}\right).
\label{spin_current_1}
\end{equation} 
The spin polarization is defined by the ratio of the spin current to the particle current $J$
\begin{equation}
 \mathcal{P}=\frac{2e}{\hbar}\frac{\mathcal{J}^y_s}{J}= \frac{\sum_{nm}\mathrm{Im}\left(t^*_{m\uparrow,n\uparrow}t_{m\downarrow,n\uparrow}\right)}{2\sum_{nm}\left(|t_{m\uparrow,n\uparrow}|^2+|t_{m\downarrow,n\uparrow}|^2\right)}.
\label{polarization}
\end{equation} 

Using explicit expressions for the transmission matrix [Eq.~(\ref{trasmis_matrix})] and the symmetry relations [Eqs~(\ref{Fxy_symmetry}) and (\ref{fmn_symmetry})] one finds 
the spin current [Eq.~(\ref{spin_current_1})] in the form
\begin{equation}
 \mathcal{J}^y_s=\sum\limits_n^N \mathcal{J}_{s,n}\,,
\label{spin_current_2}
\end{equation} 
where $\mathcal{J}^y_{s,n}$ is the partial spin current generated by electrons incident on the QPC in $n$th subband
\begin{equation}
\mathcal{J}_{s,n}\!=\!\frac{eV}{2\pi}\left(\!\frac{m^*\alpha_0L}{\hbar^2}\!\right)^3\!\frac{L^2}{w^2} \sum\limits_m \tilde f^2_{mn}D_{mn}\,, 
\label{spin_current_3}
\end{equation} 
\begin{equation}
\begin{split}
 D_{mn}=\frac{1}{q_m} &\left[\left(\frac{|t_n|^2}{q_n}+\frac{|t_m|^2}{q_m}\right)\left(A_{mn}-A_{nm}\right)\right.\\
&\left.+\frac{|t_n|^2}{q_n}\left(B_{mn}+B_{nm}\right)-\frac{|t_m|^2}{4q_m}C_{mn}\right] \,,
\end{split}
\label{D_matrix}
\end{equation} 
\begin{equation}
 A_{mn}=PV\int\limits_0^{\infty}\frac{dq}{2\pi}\frac{\mathrm{Im}\left[\tilde F^y_{+k_m,+k_n}\sum_r\tilde F^x_{+k_n,rk}\tilde F^y_{rk,+k_m}\right]}{q^2_n-q^2}\,,
\label{A_matrix}
\end{equation} 
\begin{equation}
 B_{mn}=PV\int\limits_0^{\infty}\frac{dq}{2\pi}\frac{\mathrm{Im}\left[\tilde F^y_{-k_m,+k_n}\sum_r\tilde F^x_{+k_n,rk}\tilde F^y_{rk,-k_m}\right]}{q^2_n-q^2}\,,
\label{B_matrix}
\end{equation} 
\begin{equation}
 C_{mn}\!=\!\left(\!\frac{\tilde F^x_{-k_n,+k_n}}{q_n}+\frac{\tilde F^x_{-k_m,+k_m}}{q_m}\!\right)\mathrm{Re}\!\left[\!\tilde F^y_{+k_m,+k_n}\tilde F^y_{-k_n,+k_m}\!\right],
\label{C_matrix}
\end{equation} 
where $PV$ denotes the Cauchy principal value.

Here we have gone to dimensionless (marked by tilde) variables by introducing: $q=kL$ for the wave vector, $\tilde f_{mn}=wf_{mn}$ for transverse components of the matrix elements. The function describing the spatial distribution of SOI is normalized by the amplitude of the SOI strength $\alpha_0$
\begin{equation*}
 \tilde{\alpha}(x/L_{so})=\alpha(x)/\alpha_0,
\end{equation*}
with $L_{so}$ being a characteristic length. The dimensionless components of the longitudinal matrix elements are
\begin{equation*}
\tilde F^x_{r_2k_2,r_1k_1}\!\!=\!F^x_{r_2k_2,r_1k_1}/\alpha_0,\,\, \tilde F^y_{r_2k_2,r_1k_1}\!\!=\!F^y_{r_2k_2,r_1k_1}/L\alpha_0.
\end{equation*} 

If the confinement potential is parabolic [Eq.~(\ref{potential_landscape})], the intersubband transitions are possible only for $m=n \pm 1$, according to Eq.~(\ref{fmn_parabolic}). In this case, Eq.~(\ref{spin_current_2}) for the spin current reduces to
\begin{equation}
 \mathcal{J}_{s,n}\!=\!\frac{eV}{4\pi}\left(\!\frac{m^*\alpha_0L}{\hbar^2}\!\right)^3\!\frac{L^2}{w^2}\left[nD_{n-1,n}+(n+1)D_{n+1,n}\right].
\label{spin_current_n}
\end{equation} 

Equations (\ref{spin_current_3}) and (\ref{D_matrix}) show that the spin current appears in the third order of the perturbation theory and is determined by the product of three matrix elements two of which represent intersubband transitions, and one matrix element corresponds to an intrasubband transition. These transitions include processes of both forward- and back-scattering with the change in the longitudinal wave vector.

The analysis of matrix elements $F^{x,y}_{r_2k_2,r_1k_1}$ [Eqs~(\ref{F_matrix})] shows that they all are large only when $k_1$ and $k_2$ are close to each other. The decrease in $|F^{x,y}_{r_2k_2,r_1k_1}|$ with the difference $|k_2-k_1|$ is determined by the function $\alpha (x)$. This is why the spin current is affected by the spatial distribution of the SOI strength. This effect is characterized by the parameter $|k_m-k_n|L_{so}$.

The expression for the spin current contains the factor
\begin{equation}
\mathcal{G}_{s0}=\frac{1}{4\pi}\left(\!\frac{m^*\alpha_0L}{\hbar^2}\!\right)^3\!\frac{L^2}{w^2}\,,
\end{equation}  
which largely determines the magnitude of the spin conductance. It is the product of two parameters. One parameter $(m^*\alpha_0L/\hbar^2)$ is equal to $k_{so}L$, with $k_{so}$ being the characteristic wave vector of the SOI. This parameter should not be necessarily small, since the perturbation theory requires only the characteristic energy of the SOI ($E_{so}=m^*\alpha_0^2/2\hbar^2$) to be smaller than the kinetic energy, which is on the order of the Fermi energy $E_F$ and intersubband energy $\hbar\omega_y$. Another parameter, $L/w$, has the meaning of the ratio of the length of a ``quantum wire'' formed in the QPC to its width. This is a large value for the QPCs, in which the charge conductance quantization is well pronounced.~\cite{Buttiker} Thus, the factor $\mathcal{G}_{s0}$ can be large.

Thus, the spin conductance is
\begin{equation}
 \mathcal{G}_{s}=\frac{\mathcal{J}^y_s}{eV}=\mathcal{G}_{s0} \mathcal{S}[\alpha(x)] \,,
\end{equation} 
where ${\mathcal{S}}[\alpha(x)]$ is the normalized spin conductance that depends on the spatial distribution of the SOI strength. 

In the case of the parabolic confinement potential, the spin current [Eq.~(\ref{spin_current_n})] produced by incident electrons of $n$th subband contains only two components, which correspond to the transitions through the QPC via the nearest upper and lower subbands.

The particle current $J$ and the charge conductance $G$ are easily calculated in the same way as above resulting in the following expression for $G$:
\begin{equation}
 G=\frac{2e^2}{h}\sum_n \tilde T_n,,
\label{charge_conductance}
\end{equation} 
where $\tilde T_n=\sum_{m}\left(|t_{m\uparrow,n\uparrow}|^2+|t_{m\downarrow,n\uparrow}|^2\right)$ is the transmission coefficient  modified by the SOI:
\begin{widetext}
\begin{equation}
 \tilde T_n\approx |t_n|^2\!-\!\left(\!\frac{m^*\alpha_0L}{\hbar^2}\!\right)^2\!\left\{\!\frac{|t_n|^2}{q_n^2}\left|\tilde F^x_{-k_n,+k_n}\!\right|^2\!+\!\frac{L^2}{w^2}\sum_m\tilde f_{mn}^2\!\left[\left(\!\frac{|t_n|^2}{q_nq_m}\!-\!\frac{|t_m|^2}{q_m^2}\!\right)\left|\tilde F^y_{+k_n,+k_n}\!\right|^2\!+\!\frac{|t_n|^2}{q_nq_m}\left|\tilde F^y_{-k_n,+k_n}\!\right|^2 \!\right] \!\right\}.
\label{eff_transmittance}
 \end{equation} 
\end{widetext}
Equation~(\ref{eff_transmittance}) shows that the SOI modifies the conductance in the second order of its strength. Here, the first term in the braces describes the transmittance decrease due to intrasubband backscattering. The second term containing the large factor $(L/w)^2$ corresponds to the intersubband scattering processes.

Straightforward calculations of the spin conductance were carried out for the spatial distribution of the SOI strength in the form
\begin{equation}
 \alpha (x)=\alpha_0 \exp[-x^2/L^2_{so}]
\end{equation} 
with using Eqs~(\ref{spin_current_n}) and (\ref{D_matrix}). The results are presented in Fig.~\ref{saddle_gen} for practically most interesting case where $L_{so}$ is close to $L$. Here and in the following figures, the energy is normalized to $\varepsilon_L=\pi^2\hbar^2/(2m^*L^2)$. Figure~\ref{saddle_gen} shows both the total spin conductance and the partial spin conductances defined via the spin currents $\mathcal{J}_{s,n}$ produced by electrons incident on the QPC in $n$th subband. 

\begin{figure}
\includegraphics[width=0.95\linewidth]{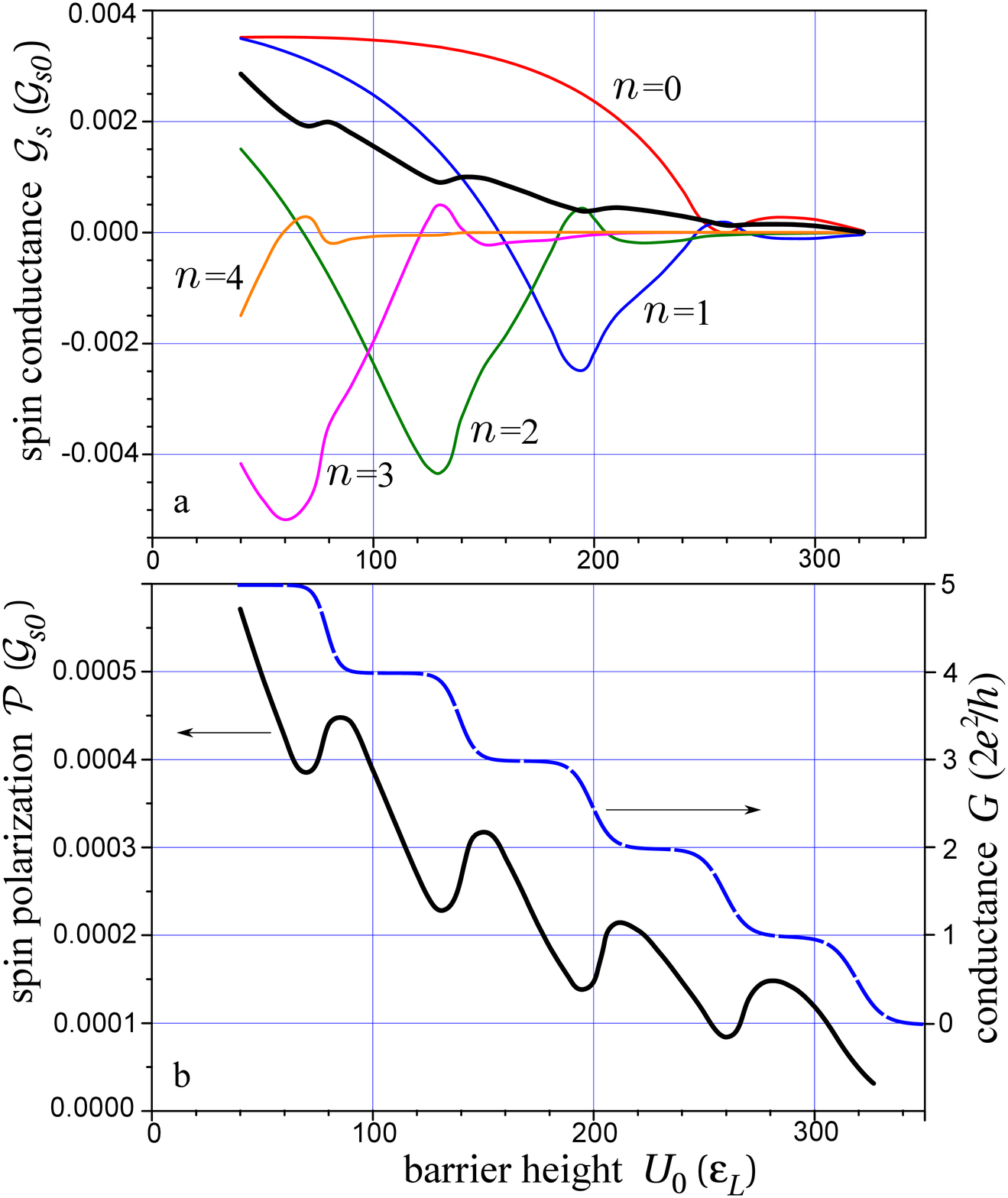}
\caption{(Color online) Spin conductance, spin polarization and charge conductance of the QPC as functions of the barrier height. (a)  The total spin conductance (thick line) and partial spin conductances caused by electrons incident on the QPC in $n$th subband (thin lines). (b) The spin polarization (full line) and the charge conductance (dashed line). The parameters used in calculations are: $E_F=350\varepsilon_L$; $\hbar\omega_y=60\varepsilon_L$; $L_{so}=L$.}
\label{saddle_gen}
\end{figure}

Consider what happens when the potential barrier $U_0$ is lowered from a large value, under which the QPC is pinched-off. When the zeroth subband ($n=0$) begins to open, the spin current arises with some delay after the electron current appears. This is because the spin current is generated by incident electrons of zeroth subband owing to their transitions to the upper ($n=1$) subband, which opens later. Transitions to the lower subband are absent in this case. With further lowering the barrier, the partial spin current $\mathcal{J}_{s,0}$ increases. 

Electrons incident on the QPC in the first subband begin to penetrate through the barrier when $U_0$ is close to the second step of the charge conductance quantization. But in contrast to the zeroth subband case, the electron transitions to the lower ($n=0$) subband are now possible and moreover, they are more effective than the transitions to the upper subband, because this subband is closed yet. Transitions to the lower subband are seen to generate the spin polarization of opposed sign. Therefore, the partial spin current $\mathcal{J}_{s,1}$ is negative from the beginning and increases in magnitude until the transitions to the upper subband become more intense. Further lowering the barrier leads to decreasing the modulus of $\mathcal{J}_{s,1}$, changing the sign and finally results in an increase in $\mathcal{J}_{s,1}$.

Spin currents produced by electrons of other upper subbands ($n= 2, 3, \dots$) behave similarly with changing $U_0$. The total spin current increases non-monotonically as the barrier is  lowered, so that the graph of the function $\mathcal{J}^y_s(U_0)$ resembles an inclined sawtooth-like line. The characteristic feature of this curve is a sequence of maxima and nearby inflection points located close to the positions of the charge conductance quantization steps.~\cite{note2}

The spin polarization defined by Eq.~(\ref{polarization}) demonstrates a similar behavior with varying $U_0$, but the extrema are much more pronounced.

To clarify how the results are changed with the size of the region, where the SOI is localized, and the barrier form we consider below two limiting cases: (i)~the SOI is strongly localized within the QPC, and (ii)~the form of the potential barrier of the QPC is sharp.

\subsection{Localized SOI strength}
Suppose that the characteristic length $L_{so}$ of the region, where the SOI acts, is much shorter than the QPC length $L$. In this case, the spin current is expressed via the longitudinal wave function and its derivative in the center of the QPC
\begin{equation}
 \psi^{(r)}_{k_n}\Big|_{x=0}\!\!=t_{k_n}F_{k_n}, \quad \frac{d}{dx}\psi^{(r)}_{k_n}\Big|_{x=0}\!\!=\!-\frac{r}{2L}t_{k_n}F_{k_n}',
\end{equation} 
where
\begin{equation}
 F_{k_n}={_2F_1}(a,b,c;\xi)\Big|_{\xi=0},\; F_{k_n}'=\frac{d}{d\xi}\left[{_2F_1}(a,b,c;\xi)\right]_{\xi=0}.
\end{equation} 

The matrix elements $F^{x,y}_{r_1k_1,r_2k_2}$ are simplified to
\begin{equation}
\begin{split}
 F^y_{r_1k_1,r_2k_2}&=\bar{\alpha}_0t^*_{k_1}t_{k_2}F^*_{k_1}F_{k_2}\\
 F^x_{r_1k_1,r_2k_2}&=\frac{\bar{\alpha}_0}{4L}t^*_{k_1}t_{k_2}(r_1{F'}^*_{k_1}F_{k_2}-r_2F^*_{k_1}F'_{k_2}),
\end{split}
\end{equation} 
where $\bar{\alpha}_0=\int_{-\infty}^{\infty}dx\alpha(x)$.

Expressions (\ref{A_matrix}), (\ref{B_matrix}), and (\ref{C_matrix}) for matrices $A_{mn}$, $B_{mn}$, and $C_{mn}$ are simplified to
\begin{equation*}
 A_{mn}=B_{mn}=\frac{1}{2}|t_{k_m}F_{k_m}|^2|t_{k_n}F_{k_n}|^2 g(q_n) \mathrm{Im}\frac{F'_{k_n}}{F_{k_n}}\,,
\end{equation*}
\begin{equation*}
\begin{split} 
C_{mn}=&-\frac{1}{2}|t_{k_m}F_{k_m}|^2|t_{k_n}F_{k_n}|^2 \\ 
&\times\left(\frac{|t_{k_n}F_{k_n}|^2}{k_n}\mathrm{Re}\frac{F'_{k_n}}{F_{k_n}}+\frac{|t_{k_m}F_{k_m}|^2}{k_m}\mathrm{Re}\frac{F'_{k_m}}{F_{k_m}}\right)\,,
\end{split}
\end{equation*} 
where $g(q_n)$ is an important function appearing from the integration over the wave vector
\begin{equation}
  g(q_n)\equiv g_n = q_n\cdot PV\int\limits_0^{\infty}\frac{dq}{2\pi}\frac{|t_{q}F_{q}|^2}{q^2-q_n^2}\,.
\end{equation}
Finally the $D_{mn}$ matrix takes the form
\begin{equation}
\begin{split}
 D_{mn}&=\frac{1}{2q_m}|t_{k_m}F_{k_m}|^2|t_{k_n}F_{k_n}|^2 \left[2\frac{|t_{k_n}|^2}{q_n^2}g_n \mathrm{Im}\frac{F'_{k_n}}{F_{k_n}} \right. \\
 &+\frac{|t_{k_m}|^2}{q_m^2}\left(g_n \mathrm{Im}\frac{F'_{k_n}}{F_{k_n}}-g_m \mathrm{Im}\frac{F'_{k_m}}{F_{k_m}}\right. \\
 &\left.\left. + \frac{|t_{k_n}F_{k_n}|^2}{4q_n}\mathrm{Re}\frac{F'_{k_n}}{F_{k_n}}+\frac{|t_{k_m}F_{k_m}|^2}{4q_m}\mathrm{Re}\frac{F'_{k_m}}{F_{k_m}}\right)\right].        
\end{split}
\label{Dmn_local}   
\end{equation} 

Equation (\ref{Dmn_local}) together with Eqs~(\ref{spin_current_2}) and (\ref{spin_current_3}) give the spin conductance as a function of the barrier height and the Fermi energy. 

In the case of parabolic confining potential, the spin conductance is expressed through the matrix elements for nearest subband transitions $D_{n\pm 1,n}$ according to Eq.~(\ref{spin_current_n}). 

\begin{figure}
\includegraphics[width=0.95\linewidth]{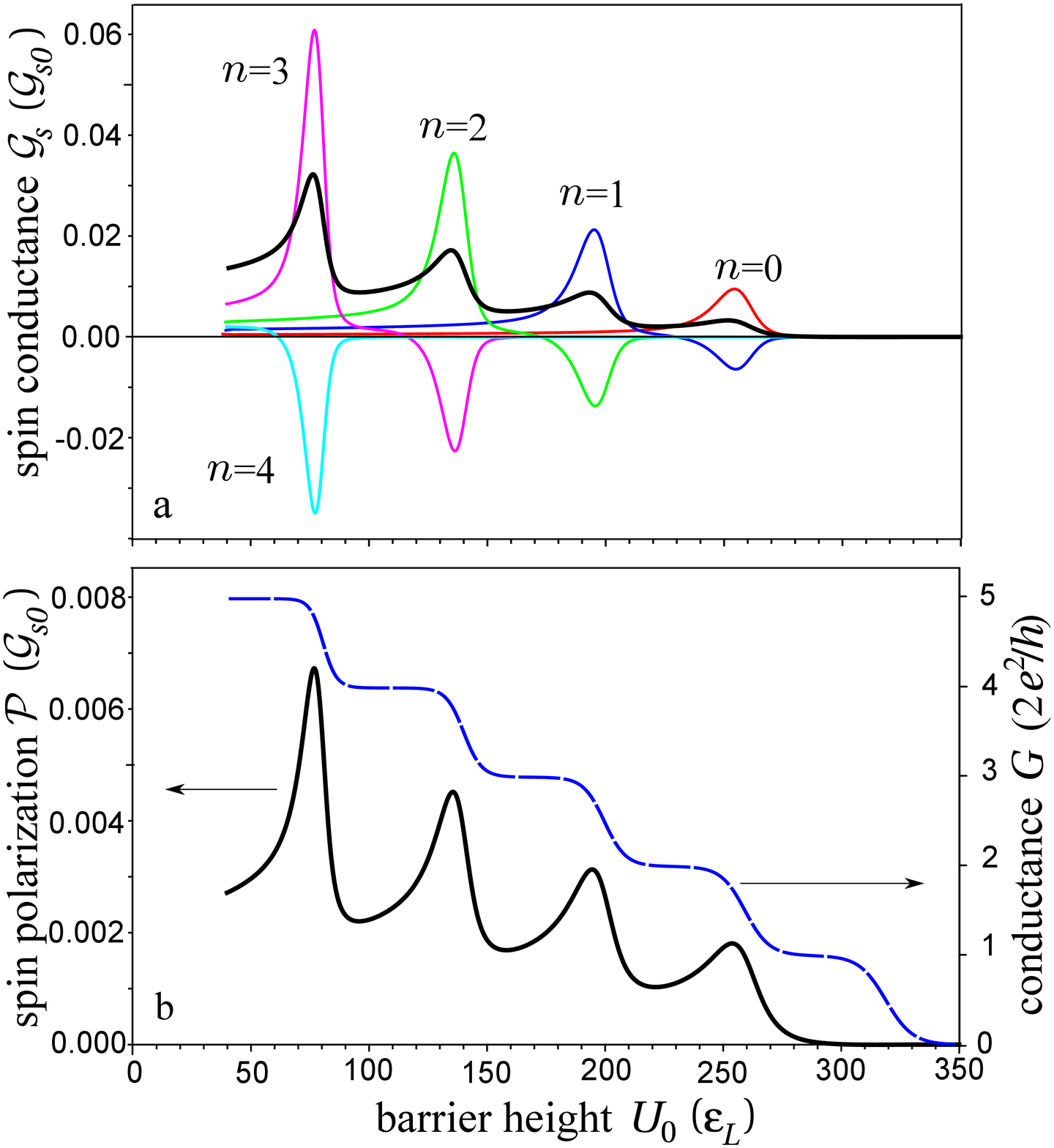}
\caption{(Color online) Spin conductance, spin polarization and charge conductance of the QPC as functions of the barrier height for the case where the SOI is localized in the center of the QPC. Captions to figures (a) and (b) are the same as in Fig.~\ref{saddle_gen}.}  
\label{saddle_local}
\end{figure}

In Fig.~\ref{saddle_local} we present the spin conductance and spin polarization calculated for the same saddle-point potential as used in Fig.~\ref{saddle_gen}, where the SOI is distributed over wide region. The only difference is that the SOI is strongly localized in the center of the QPC. The spin polarization and spin conductance are seen to behave generally in the same manner in both cases. However, in the case of the localized SOI the peaks are much more sharp.

This fact can be understood taking into account that in this case the spin conductance depends only on the local value of the wave function amplitude in the center of the QPC $|\psi^{(r)}_{k_n}(0)|=|t_{k_n}F_{k_n}|$, in contrast to the case of distributed SOI, where the wave function is integrated over the length on the order of $L_{so}$. Therefore, in order to clarify the nature of the spin-conductance peaks it is needed to know how the wave function in the center of the QPC changes with the barrier height.

\begin{figure}
\includegraphics[width=0.95\linewidth]{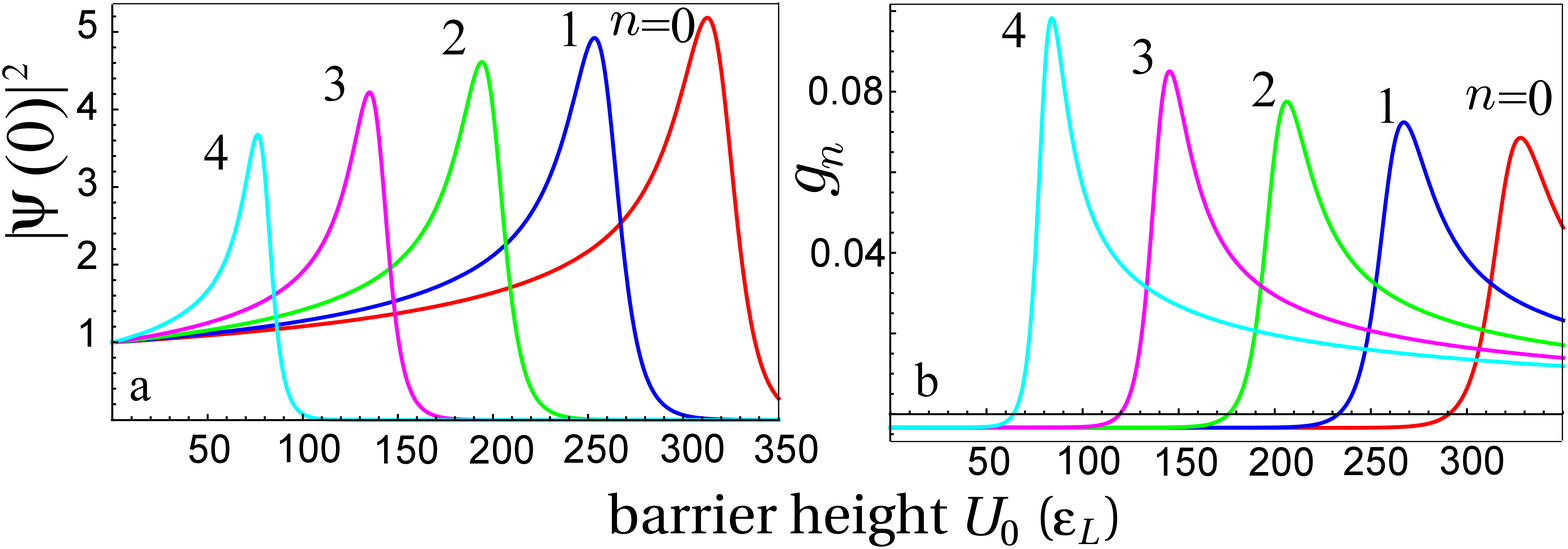}
\caption{(Color online) (a) The wave function amplitudes $|t_{k_n}F_{k_n}|^2$ and (b) the functions $g_n$ versus the barrier height for subbands $n=0, 1, 2, 3, 4$. The parameters used in the calculations are the same as in Fig.~\ref{saddle_gen}.}  
\label{wave_func_saddle}
\end{figure}

The answer to this question contains in Fig.~\ref{wave_func_saddle}. The wave function has a quite sharp peak which is reached when the barrier height is close to the electron energy. The peak is interpreted as follows. The wave function increases with lowering the barrier since the tunneling probability increases, until $E<U_0$. But as $E>U_0$, the wave function decreases because the electron velocity grows. The function $g_n(U_0)$ also has a peak. The maxima of both functions are seen to be located practically at the same positions as the spin conductance peaks. 

The above explanation allows one to suppose that the spin polarization peaks in Fig.~\ref{saddle_gen} (the case of $L\sim L_{so}$) has the same nature. The smoother shape of the peaks here is caused by the fact that in this case the spin conductance is determined by an integral of $\psi^{(r)}_{k_n}$ over the region, where the SOI is localized.

\subsection{A rectangular barrier}
In this section we consider a QPC with sharp potential landscape to study how the multiple reflections of electron waves and the interference distort the spin conductance and its dependence on the barrier height. Let the barrier be rectangular:
\begin{equation*}
 U(x,y)=U_0\left[\Theta\left(\!x\!-\!\frac{L}{2}\right)-\Theta\left(\!x\!+\!\frac{L}{2}\right)\right]+\frac{m^*\omega_y^2y^2}{2}\,,
\end{equation*} 
and the spatial distribution of the SOI strength be exponential:
\begin{equation*}
 \alpha(x)=\alpha_0\exp(-\kappa |x|)\,.
\end{equation*} 

\begin{figure}
\includegraphics[width=0.95\linewidth]{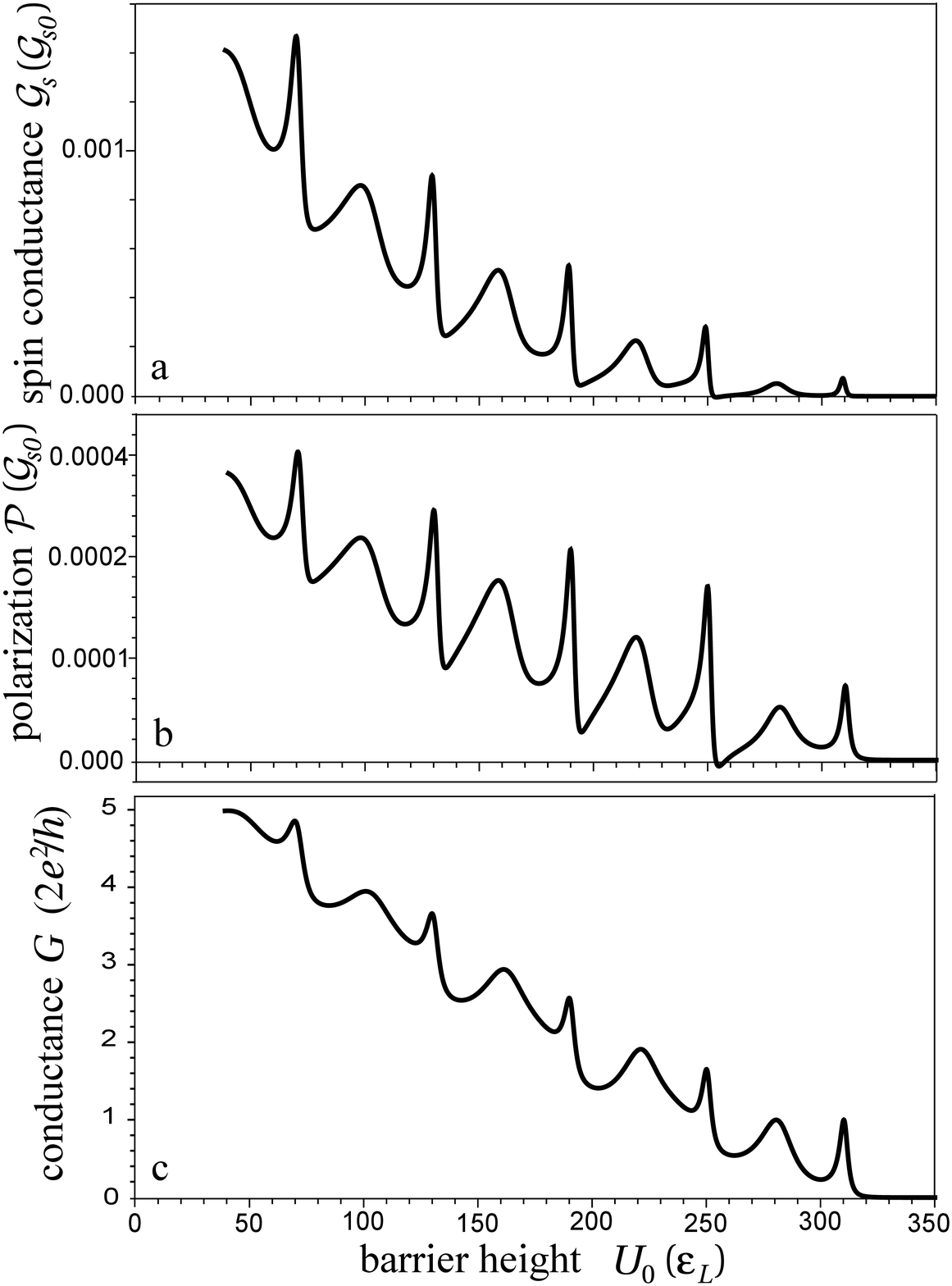}
\caption{Spin conductance, spin polarization and conductance of the QPC with rectangular barrier as functions of the barrier height. (a) The total spin conductance. (b) The spin polarization. (c) The charge conductance. The parameters used: $E_F=350\varepsilon_L$; $\hbar\omega_y=60\varepsilon_L$; $\kappa L=1$.}
\label{rectangular_1}
\end{figure}

\begin{figure}
\includegraphics[width=0.95\linewidth]{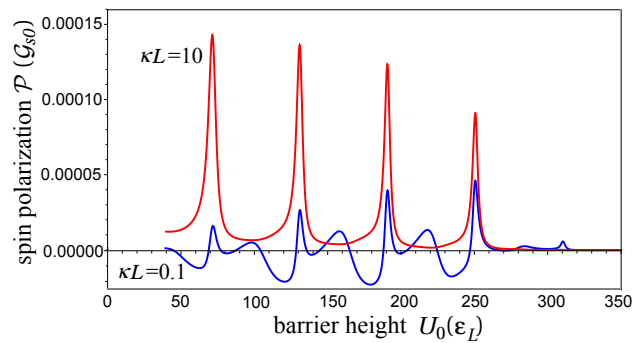}
\caption{(Color online) Spin polarization of the current through the rectangular QPC for various length of the SOI localization: strong localization ($\kappa L=10$) and weak localization ($\kappa L=0.1$). Other parameters are as in Fig.~\ref{rectangular_1}.}
\label{rectangular_2}
\end{figure}

In this case, the longitudinal wave functions are simply combinations of exponential functions, and all integrals in Eqs~(\ref{F_matrix}), (\ref{A_matrix}), (\ref{B_matrix}), and (\ref{C_matrix}) are calculated in elementary functions. Final results are illustrated in Figs~\ref{rectangular_1} and \ref{rectangular_2} for different lengths of the SOI region.

Figure~\ref{rectangular_1} shows the spin conductance and the spin polarization as functions of the barrier height $U_0$ for the case where $\kappa L=1$. The charge conductance is also plotted here. It makes sense to compare these graphs with those of Fig.~\ref{saddle_gen} since all parameters used in their calculations have been chosen the same, with the exception of the barrier form: the longitudinal barrier is rectangular rather than smooth [of $\cosh^{-2}(x/L)$ form]. The interference of reflected waves is seen to strongly distort the charge conductance plateaus. Similarly, the graphs of the spin conductance and spin polarization are also defaced and only a general trend of the curves (i.e., a shape after some averaging of the interference pattern) resembles that of corresponding curves in the case of the smooth QPC.

If the SOI strength is strongly localized within the QPC, the interference impact on the spin polarization becomes not such destructive.  Figure~\ref{rectangular_2} shows that when $\kappa L=10$, the spin-polarization curve much more resembles the corresponding graph for the saddle-point QPC, Fig.~\ref{saddle_local}. On the contrary, if the SOI acts in the wide region ($\kappa L=0.1$) the interference effect enhances. The spin polarization strongly oscillates with $U_0$ and even changes the sign under certain conditions. Such a behavior is obviously caused by complex spin dynamics occurring under the conditions of the multiple reflectance and interference. 

\section{Concluding remarks}
We have studied the spin current in QPCs with SOI using the model that considers the QPC as a saddle point of two-dimensional potential landscape with the SOI localized in a finite region. This approach has allowed us to avoid the complicated spin dynamics caused by multiple reflections and interference of electron waves and to study the spin-polarization mechanism associated with electron transitions in QPCs. Within this model we have found out quite general features of the spin conductance quantization correlated with the well-known staircase of the charge conductance.

It is shown that the spin polarization arises as a result of intersubband and intrasubband electron transitions. We have taken into account all possible transitions, not only those which were discussed by Eto et al.~\cite{Eto} It turns out that of most efficiency are definite combinations of three transitions: two intersubband transitions (one transition from a given subband to the nearest subband and a backward transition) and one intrasubband transition. We clarify that transitions via the upper and lower subbands produce the spin polarization of opposed signs. More specifically, the electron transitions to the upper subband and the following backward transitions give rise to a positive polarization, whereas transitions via the lower subband produce a negative polarization.

Within the perturbation theory, the spin polarization has a cubic dependence on the SOI strength. This result can be interpreted as follows. It is obvious that the polarization should be an odd function of the SOI strength, however, the first-order correction in the SOI is not enough to create spin polarization. In fact, the first-order correction to an incoming wave function $\psi_k(x)\varphi_n(y)|s\rangle$ is the sum of functions $\psi_{k'}(x)\varphi_m(y)|s'\rangle$. Since the transverse wave functions $\varphi_n(y)$ and $\varphi_m(y)$ of different subbands are orthogonal, the first-order correction to the spin density and the spin current could be nonzero only when $m=n$, i.e., in the case when electrons remain in the same subband after the scattering . However, it is well known that an electron flow passing through a SOI region within one subband does not acquire any spin polarization.~\cite{Governale} The point is that, though the electrons acquire a spin polarization when enter into the SOI region, they lose it completely after the exit. Therefore, only the third-order correction to the spin current can be nonzero. This our conclusion agrees with the cubic dependence of the spin polarization on the Rashba coupling strength reported in Ref.~\onlinecite{Tripathi} for uniform multimode quantum wire.

In contrast to the spin conductance, the charge conductance is an even function of the SOI strength as Eqs~(\ref{charge_conductance}) and (\ref{eff_transmittance}) show.

The spin current magnitude is determined by two factors. One factor is connected with the geometrical sizes of the saddle-point potential and the characteristic wave vector of SOI: $(k_{so}L)^3(L/w)^2$. It allows one to roughly estimate the dependence of the spin conductance on $L$ and $w$. Other factor is connected with the shape of the spatial distribution of the SOI strength and the form of the potential barrier in the QPC. It is this factor that determines the dependence of the spin polarization and spin conductance on the barrier height.

Note, that in the presence of the SOI, the dependences of the spin conductance on barrier height $U_0$ and the Fermi energy $E_F$ are qualitatively different, in contrast to the case without the SOI, where they are similar.

The general feature of the spin polarization as a function of $U_0$ is the presence of peaks and nearby inflection points located close to the charge conductance quantization steps. We argue that these peaks originate from the presence of the sharp maximum that the wave-function amplitude reaches in the QPC when the barrier height is varied.

In the case where the potential landscape of the QPC is sharp, the interference of electron waves distorts this feature of the spin conductance as well as the form of the quantization plateaus of the charge conductance. The studies presented in this paper were carried out with using the approximations which restricted the applicability of the results in the following points. 

(i) The perturbation theory imposes a restriction on the SOI strength. In general form this restriction in rather complicated because perturbing Hamiltonian contains both $\alpha$ and $d\alpha/dx$. However, in the practically interesting case where the transmission coefficient in a given subband is not too small, the following rough estimation is obtained
\begin{equation}
 \frac{\bar \alpha_0\hbar^2}{m^*}\ll \sqrt{\frac{E}{\hbar\omega_y}\left(\frac{1}{1+w\bar k}\right)}\,,
\label{restrict}
\end{equation} 
where $\bar \alpha_0=\int dx\alpha (x)$, $\hbar\omega_y$ is the transverse quantization energy, $\bar k$ is characteristic wave vector of electrons in the QPQ. The inequality [Eq.~(\ref{restrict})] is obtained for $|t_n|> \bar \alpha_0\hbar^2/m^*$. Under realistic conditions of experiments, $w\bar k\sim 1$, $E/\hbar\omega_y\sim 1$. With using the SOI parameter $\alpha_0$ for InAs and GaAs, one sees that the restriction [Eq.~(\ref{restrict})] is well satisfied.

(ii) The conclusions about spin current dependences on the SOI strength and the barrier height $U_0$ pertain to the case where the variables $x$ and $y$ are separated in the unperturbed Hamiltonian, particularly to the case of the adiabatic separation of variables.~\cite{Glazman} If variables are not separated, the effect of spin polarization can appear in the first approximation in the SOI strength, but the spin current depends on the nonadiabaticity parameter. Note that in this case the subbands are mixed even in the unperturbed state and therefore the conductance quantization plateaus are destroyed. The question of the spin polarization requires a separate study which will be published elsewhere.

(iii) Important restriction consists also in ignoring the electron-electron interaction which is very essential at low density of electrons. In QPCs the electron-electron interaction is known to produce a strong change in the electron structure of the ground state and the electron transport in the regime where the conductance is lower than $2e^2/h$ and transport anomalies are observed, such as ``0.7 feature'' and zero-bias anomaly.~\cite{Pepper} Their nature is not clarified to date, but the most probable origin is the appearance of a spin-polarized electron state. Though our consideration is not applicable to this regime, one can expect that the SOI can give rise to a strong effect under such conditions.

\acknowledgments
This work was supported by Russian Foundation for Basic Research (Project No.~08-02-00777) and Russian Academy of Sciences (Program No.~27 ``Basic researches in nanotechnology and nanomaterials'' and program ``Strongly correlated electrons in solids and structures'').


\begin{thebibliography}{30}
\bibitem{note1}
The assumption that the SOI vanishes in the reservoirs is widely used in the literature to find the spin polarization acquired by an electron current flowing through a mesoscopic system with SOI. It is enough to refer to above Refs~\onlinecite{ZhaiPRL,Reynoso1,Liu,Zhai2,Entin-Wohlman}.
\bibitem{note2}
Similar features of the curves describing the spin conductance versus $U_0$ were obtained in numerical calculations of Ref.~\onlinecite{Reynoso1}.
\end{thebibliography}
\end{document}